\documentclass[journal]{IEEEtran}
\usepackage[table]{xcolor}
\definecolor{subsectioncolor}{rgb}{0.067,0.627,0.859}
\definecolor{abstractbg}{RGB}{245,250,255}

\usepackage{epstopdf}
\usepackage{amsmath}
\usepackage{amssymb}
\usepackage[T1]{fontenc}
\usepackage{lmodern}


\def\ps@IEEEtitlepagestyle{%
  \def\@oddhead{}\def\@evenhead{}%
  \def\@oddfoot{\hfil\thepage\hfil}%
  \def\@evenfoot{\hfil\thepage\hfil}}
\makeatother

\usepackage{cite}

\usepackage{enumitem}
\usepackage{array}
\usepackage{makecell}

\newcolumntype{P}[1]{>{\centering\arraybackslash}m{#1}}
\usepackage{amsmath,amssymb,amsfonts}
\usepackage{algorithmic}
\usepackage{graphicx}
\DeclareGraphicsExtensions{.pdf,.png,.jpg,.eps}
\usepackage{textcomp}
\usepackage[utf8]{inputenc}
\usepackage{newunicodechar}
\newunicodechar{₃}{$_3$}
\usepackage{wrapfig,colortbl}
\usepackage{blindtext}
\usepackage{lipsum}
\definecolor{abstractbg}{rgb}{1,0.969,0.914}
\def\BibTeX{{\rm B\kern-.05em{\sc i\kern-.025em b}\kern-.08em
    T\kern-.1667em\lower.7ex\hbox{E}\kern-.125emX}}

\begin{document}
\title{Periodically Poled Piezoelectric Lithium Niobate Resonator for Piezoelectric Power Conversion  }
\author{Ziqian Yao, Clarissa Daniel, Lezli Matto, Heather Chang, Vakhtang Chuluhadze, Michael Liao, \\Jack Kramer, Eric Stolt, Mark S. Goorsky, Juan Rivas-Davila, and Ruochen Lu
\thanks{This work was supported by Defense Advanced Research Projects Agency (DARPA) High Operational Temperature Sensors (HOTS) and DARPA Transducers for Optimized Robust Nonlinear Actuation and Dynamic Operation with Speed (TORNADO) programs. }
\thanks{Z. Yao, V. Chulukhadze, J. Kramer, and R. Lu are with the Electrical and Computer Engineering Department, The University of Texas at Austin, Austin, TX, 78712 USA (email: hanson.yao@utexas.edu).}
\thanks{C. Daniel, H. Chang, E. Stolt, and J. Rivas are with the Electrical Engineering Department, Stanford University, Stanford, CA 94305 USA.}
\thanks{L. Matto, M. Liao, and M. S. Goorsky are with the University of California, Los Angeles, CA, 90095 USA.}}

\IEEEtitleabstractindextext{%
\fcolorbox{abstractbg}{abstractbg}{%
\begin{minipage}{\textwidth}\rightskip2em\leftskip\rightskip\bigskip
\begin{wrapfigure}[15]{r}{3in}%
\hspace{-3pc}\includegraphics[width=2.9in]{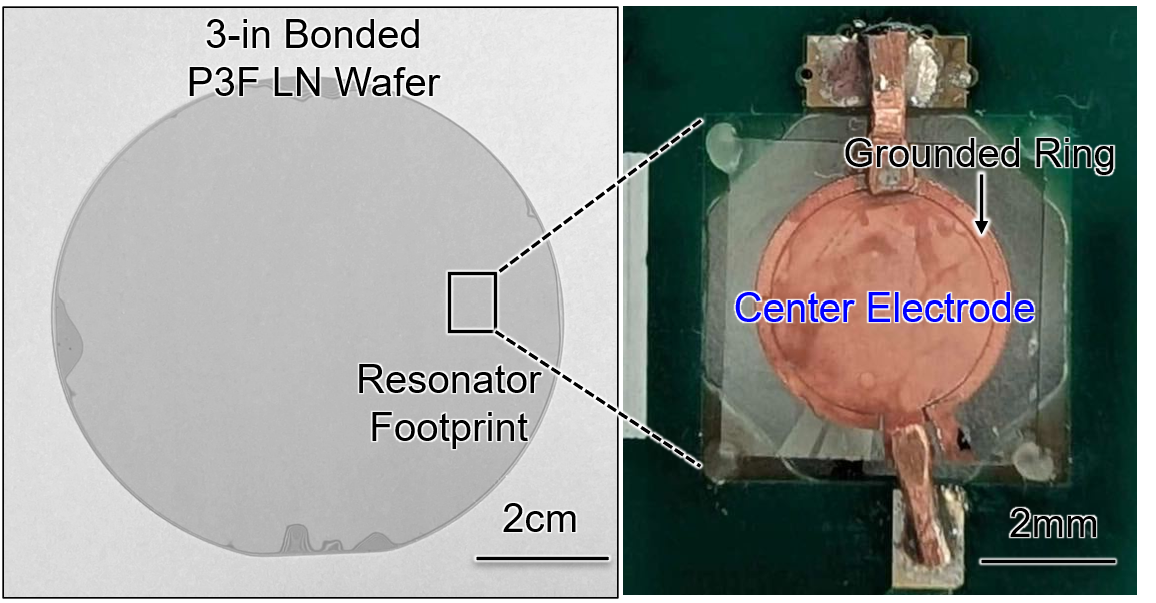}
\end{wrapfigure}%

\bigskip
\end{minipage}}}

\maketitle

\begin{abstract}
As the demand for compact and efficient power conversion systems increases, piezoelectric power converters have gained attention for their ability to replace bulky magnetic inductors with acoustic resonators, enabling higher power density and improved efficiency. Achieving optimal converter performance requires resonators with high quality factor ($Q$), strong electromechanical coupling ($k^2$), high power handling capability, and a spurious-free response. Lithium niobate (LN) has emerged as a promising material in this context due to its high figure of merit (FoM = $Q \cdot k^2$). While previous studies on single-layer LN resonators have demonstrated high FoM values, they typically operate at relatively low resonance frequencies ($f_s$). 
Recently, periodically poled piezoelectric film (P3F) structures, formed by stacking piezoelectric layers with alternating crystal orientations, have shown the potential to both scale up the operating frequency and enhance the FoM compared to single-layer counterparts in piezoelectric power conversion. This work presents the first P3F thickness-extensional (TE) LN resonator for power conversion, operating at 19.23 MHz, with a large \textit{$k^2$} of 29\% and a high \textit{Q} of 3187, achieving a state-of-the-art (\textit{ $f_s \cdot Q$}) product among piezoelectric power resonators. A high-power testing procedure is performed to systematically study the nonlinear behavior and power handling of P3F LN for power applications. With further optimization, P3F TE resonators have the potential to open up a new design space for high-power and high-frequency power conversion.
\end{abstract}

\begin{table*}[!t]
\arrayrulecolor{subsectioncolor}
\setlength{\arrayrulewidth}{1pt}
{\sffamily\bfseries\begin{tabular}{lp{6.75in}}\hline
\rowcolor{abstractbg}\multicolumn{2}{l}{\color{subsectioncolor}{\itshape
Highlights}{\Huge\strut}}\\
\rowcolor{abstractbg}$\bullet${\large\strut} & A bi-layer lithium niobate resonator is demonstrated using a periodically poled piezoelectric film (P3F) structure to enable second-order antisymmetric (A2) thickness-extensional mode excitation for power conversion.\\
\rowcolor{abstractbg}$\bullet$ {\large\strut} & The fabricated resonator demonstrates high $f$ of 19.23 MHz, $k^2$ of 29\%, and $Q$ of 3187, yielding a state-of-the-art $f \cdot Q$ product among piezoelectric power resonators. The P3F structure enables higher-order mode excitation without sacrificing $Q$, offering a scalable platform for high-frequency, high-power piezoelectric power conversion \\
\rowcolor{abstractbg}$\bullet${\large\strut} & A systematic study of nonlinear behavior and power handling testing is reported up to 58 dBm. \\[2em]\hline
\end{tabular}}
\setlength{\arrayrulewidth}{0.4pt}
\arrayrulecolor{black}
\end{table*}

\section{Introduction}
\label{sec:introduction}

\IEEEPARstart{P}{ower} converters demand compact, high-efficiency operation with high power density. However, magnetic-based inductors, commonly used for passive energy storage, present scalability challenges due to the inherently large wavelengths of electromagnetic (EM) waves and increased parasitic effects at high switching frequencies \cite{sullivan2016size,zhao2022comparative, hanson2016measurements}. These limitations hinder efforts to reduce converter size while maintaining performance. To overcome these challenges, piezoelectric acoustic resonators offer a promising alternative by converting EM energy into mechanical vibrations, enabling smaller footprints and lower switching losses due to shorter wavelengths\cite{ruby2015snapshot, gong2021microwave, hassanien2020acoustically}. 
When operated near their series resonance frequency, these resonators enable resonant energy transfer by soft-charging their static capacitance \cite{piel2021feedback,braun2021optimized, boless2022piezoelectric, forrester2022resonant}. Recent work on materials such as lithium niobate (LN) \cite{stolt2024stacked,braun2024stacked,stolt2023spurious}, lithium tantalate (LT)\cite{yao2025lithium} and lead zirconate titanate (PZT) \cite{naval2025high,skinner2024piezoelectric, boles2022piezoelectric} has enabled compact, high-efficiency piezoelectric resonators that exhibits the potential to replace bulky inductors in conventional power converters \cite{kyaw2017fundamental, boles2020analysis, hou2023vertically}.

Optimal converter performance requires resonators with high quality factor ($Q$), strong electromechanical coupling ($k^2$), high power handling, and minimal spurious modes. Although PZT radial and thickness-extensional (TE) mode resonators offer relatively high $k^2$, their low resonance frequencies ($f_s$) and modest figure-of-merit (FoM = $Q \cdot k^2$) limit device scalability \cite{boles2021evaluating}. LN radial-mode devices provide higher $Q$ and FoM but remain constrained to sub-MHz operation \cite{stolt2021fixed}. Single-layer LN resonators, including thickness-shear (TS) and TE modes, offer strong $k^2$ and high $Q$ \cite{nguyen2023spurious, nguyen2023near, touhami2021piezoelectric}, making them promising candidates for compact, high-efficiency designs.

However, using thin piezoelectric layers to achieve future reduction in resonator footprint while preserving high $Q$ and strong power handling remains a key challenge \cite{schaefer2019process}. Thin piezoelectric layers are commonly used to maintain capacitance density and increase the resonance frequency. Minimizing the lateral footprint without increasing wafer thickness inherently limits power handling, as the maximum stored energy is restricted by the active volume and material limits \cite{daniel2024nonlinear}. In contrast, increasing wafer thickness improves energy storage capacity and power density but lowers the resonance frequency. Parallel-connected resonator arrays have been proposed to boost power handling \cite{stolt2024stacked}, but the necessary spacing between individual devices increases the overall footprint. These limitations underscore the need for a novel structural approach that can simultaneously reduce footprint, maintain high resonance frequency, and support robust power handling.
\begin{table}[!t]
\caption{SOA of Piezoelectric Resonator Performance}
\label{Talble:1}
\centering
\setlength{\tabcolsep}{6pt}
\renewcommand{\arraystretch}{1.3}
\begin{tabular}{P{55pt} P{20pt} P{20pt} P{20pt} P{25pt} P{35pt}}
\hline
\hline
{Reference} & $f_s$ {(MHz)} & $k^2$ & {Q} & {FoM} ($Q\cdot k^2$) & $f_s\cdot Q$ \textbf{($\times 10^{10}$)} \\
\hline
PZT-Radial\cite{boles2022piezoelectric} & 0.48  & 0.19 & 1030  & 196   & 0.05 \\
PZT-TE\cite{boles2021evaluating}       & 0.61  & 0.31 & 2500  & 775   & 0.15 \\
LN-Radial\cite{stolt2021fixed}         & 0.30  & 0.23 & 9300  & 2139  & 0.28 \\
LN-TS\cite{nguyen2023near}             & 5.94  & 0.45 & 3500  & 1575  & 2.08 \\
LN-TE\cite{touhami2021piezoelectric}   & 6.28  & 0.26 & 3700  & 944   & 2.32 \\
LN-TE\cite{nguyen2023spurious}         & 10.14 & 0.30 & 4000  & 1200  & 4.06 \\
\textbf{P3F-LN-TE (This work)}         & \textbf{19.23} & \textbf{0.29} & \textbf{3187} & \textbf{928} & \textbf{6.13} \\
\hline
\hline
\end{tabular}
\end{table}
To address these limitations, we introduce the periodically poled piezoelectric film (P3F) structure, consisting of stacked piezoelectric layers with alternating crystal orientations\cite{lu2020enabling,kramer2024experimental,barrera2023fundamental,lu2025recent}. The P3F structure can be conceptualized as a vertically stacked array of piezoelectric resonators, effectively reducing device footprint while preserving power handling capability and overall performance. Originally developed for microwave and millimeter-wave acoustic applications, P3F structures have demonstrated strong performance in wideband filters and resonators operating in the GHz regime \cite{vetury2023manufacturable,kramer2023thin,cho202423,peng202456}. Unlike conventional single-layer platforms, where scaling to higher frequencies often leads to reduced $k^2$ and degraded energy confinement due to longitudinal film thinning or lateral active-region reduction \cite{nguyen2019impact}, P3F enables operation using thicker film stacks, thereby maintaining strong acoustic energy confinement and potentially higher energy density \cite{lu2020enabling}. Furthermore, this structural flexibility empowers the extension to multilayer configurations, allowing for additional frequency scaling and bandwidth control. Nonetheless, the P3F LN structures have not been studied for high-power applications, e.g., piezoelectric power conversion.

In this work, we present the first bi-layer TE mode LN resonator using a P3F structure for power conversion. The device achieves a high resonance frequency of 19.23 MHz, $k^2$ of 29\%, and $Q$ of 3187, yielding a state-of-the-art frequency–quality factor product ($f \cdot Q$) among piezoelectric power resonators (Table\ref{Talble:1}). The details of wafer bonding, as well as the high-power testing, are presented. This article is organized as follows. Section II describes the resonator design, including crystal orientation and simulation methodology. Section III outlines the wafer bonding process and material characterization. Section IV summarizes the fabrication steps for the resonator. Section V presents the electrical measurement results, and Section VI discusses the high-power testing, including nonlinear behavior and power handling analysis.

\section{Design and simulation}
This section outlines the material orientation and structural design choices for the proposed P3F power resonators, which are validated through finite element simulations of mode shapes, stress distributions, and impedance responses.

\subsection{Orientation Consideration}

To enable efficient excitation of acoustic modes for power conversion, the choice of crystal orientation is critical. The TE resonator was designed using \text{36$^\circ$ Y} LN for its high piezoelectric coefficient \( e_{33} \) for thickness electrical field excited (\( E_z \)) TE modes (\( T_{zz}\) ) \cite{kramer202257,lu2020thin,wu2023ultra,liu20227,nguyen2023spurious}. The piezoelectric matrix for \text{36$^\circ$ Y} is listed below:

\begin{equation}
\mathbf{e}_{\text{36$^\circ$Y}} =
\resizebox{0.75\columnwidth}{!}{$
\begin{bmatrix}
0 & 0 & 0 & 0 & 0.244 & -4.470 \\
-1.614 & -2.377 & 2.543 & 0.419 & 0 & 0 \\
-1.831 & -1.655 & 4.538 & -0.313 & 0 & 0
\end{bmatrix}
$}
\label{eq:e36}
\end{equation}

In our P3F structure (Fig.~\ref{fig1}), both the top and bottom layers are cut at 36$^\circ$Y. The top layer retains its native orientation and follows the piezoelectric matrix in \eqref{eq:e36}. The bottom layer, however, is rotated in-plane by 180$^\circ$ with respect to the top one, effectively flipping the signs of piezoelectric coefficients that are odd under inversion symmetry, such as \textit{\( e_{33} \)}, \( e_{31} \), and \( e_{32} \). This transformation results in the rotated matrix shown in \eqref{eq:e36_rotated}, leading to alternating piezoelectric polarity across the film stack. 

\begin{equation}
\mathbf{e'}_{\text{36$^\circ$Y}} =
\resizebox{0.75\columnwidth}{!}{$
\begin{bmatrix}
0 & 0 & 0 & 0 & -0.244 & 4.470 \\
1.614 & 2.377 & -2.543 & -0.419 & 0 & 0 \\
1.831 & 1.655 & -4.538 & 0.313 & 0 & 0
\end{bmatrix}
$}
\label{eq:e36_rotated}
\end{equation}

It can be seen that the \( e_{31} \), \( e_{32} \), and \( e_{33} \) components in the top and bottom layers have the opposite sign. As a result, for the electrode configuration in Fig.~\ref{fig1} where the thickness electrical field across both layers is applied, both the longitudinal extensional modes will be fully cancelled, suppressing lateral-mode excitation. 

When TE modes are excited, this polarity alternation enables constructive charge buildup in symmetric even-order acoustic modes (e.g., second-order antisymmetric, A2), which is the second-order TE mode. In contrast, such overtones are typically suppressed in single-layer LN due to charge cancellation from opposing stress-phase regions across the thickness. By restoring coupling in these modes and supporting thicker film stacks, the P3F platform breaks the conventional trade-off between frequency scaling and coupling efficiency, achieving high power density, high \( k^2 \), high $Q$, and high FoM simultaneously.

\subsection{Device Design and Simulation}

\begin{figure}[t!]
\centerline{\includegraphics[width=\columnwidth]{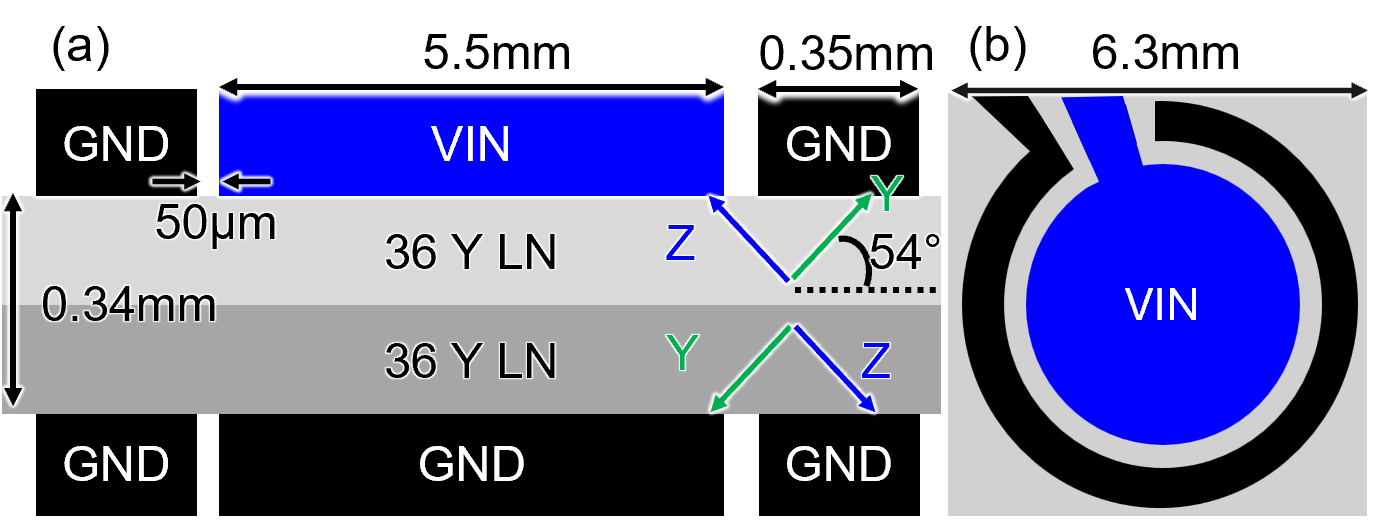}}
\caption{Schematics of (a) side-view with material axes of P3F structure, where the two LN layers are oppositely poled, and (b) top-down view.}
\label{fig1}
\end{figure}

The proposed P3F TE-mode resonator is shown in Fig.~\ref{fig1}. It consists of circular top and bottom electrodes patterned on a stacked 0.34 mm thick 36$^\circ$Y P3F LN wafer, targeting an operating frequency near 19.5 MHz. The top electrode features a 5.5 mm wide central VIN region, surrounded by a 0.35 mm grounded ring with a 50~µm separation gap. The bottom side mirrors this layout with segmented grounded areas aligned to the top pattern. This grounded ring structure is adapted from the single-layer TE-mode resonator design in \cite{nguyen2023spurious}, which used a 0.3 mm-thick LN wafer. While the grounded ring effectively suppresses in-band spurious modes in the single-layer case, its suppression effect is less pronounced in the P3F configuration. This may be attributed to the opposite group velocity directions of S1 and A2 Lamb modes, which alter lateral wave propagation and reduce the ring's ability to reflect or dissipate in-band spurious energy. Although further study is required to quantify this effect fully, we retain the same electrode layout to enable a direct and fair comparison between the P3F and single-layer designs.

\begin{figure}[!t]
\centerline{\includegraphics[width=\columnwidth]{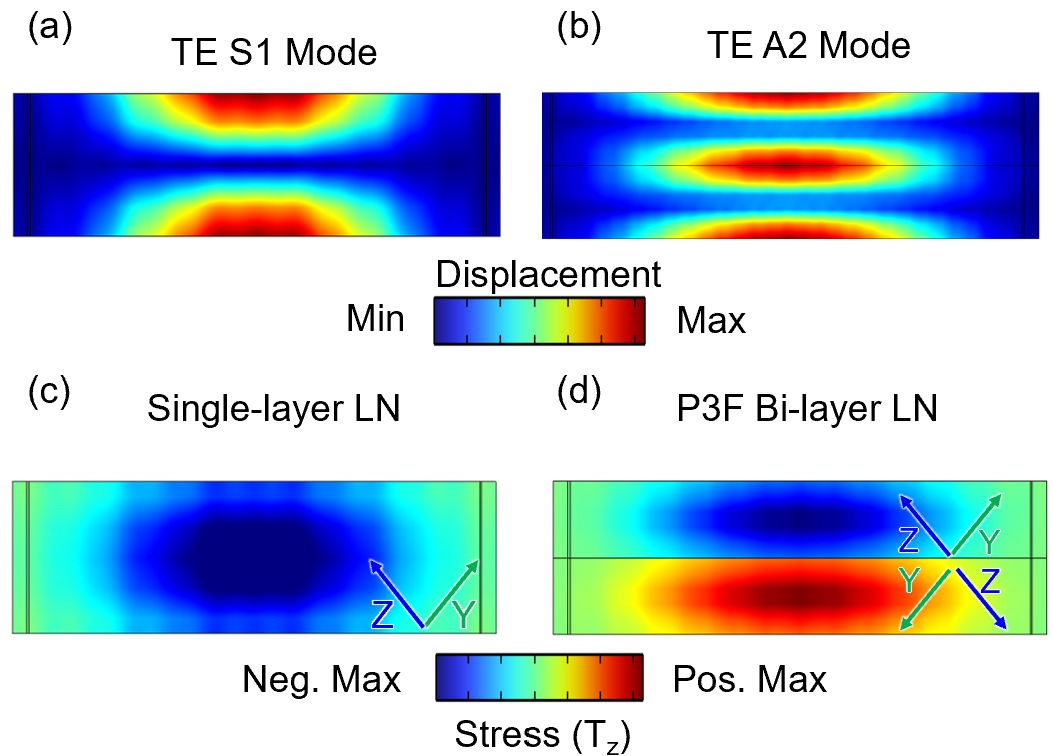}}
\caption{Simulated mode shapes of single-layer and P3F bi-layer LN:(a),(b) Displacement and stress of single-layer LN; (c), (d) Displacement and stress of P3F bi-layer LN.}
\label{fig2}
\end{figure}

 To understand the physical mechanism enabling A2 excitation in the P3F design, we use finite element analysis (FEA) in COMSOL Multiphysics to compare simulated mode shapes between a single-layer LN and a P3F bi-layer structure, as shown in Fig.~\ref{fig2}. In the single-layer case, the TE S1 mode Fig.~\ref{fig2}(a) dominates due to suppression of even-order overtones caused by charge cancellation between opposing stress-phase regions. The corresponding stress distribution Fig.~\ref{fig2}(c) confirms this symmetry and limited excitation of even modes. In contrast, the P3F bi-layer configuration supports strong excitation of the TE A2 mode, Fig.~\ref{fig2}(b), enabled by alternating crystal polarity, which promotes constructive charge buildup for symmetric even-order modes. The resulting stress field Fig.~\ref{fig2}(d) shows complementary compressive and tensile regions across the stack, validating the P3F design's effectiveness in restoring coupling to higher-order modes\cite{ruby2007recessed}.
 
\begin{figure}[!t]
\centerline{\includegraphics[width=\columnwidth]{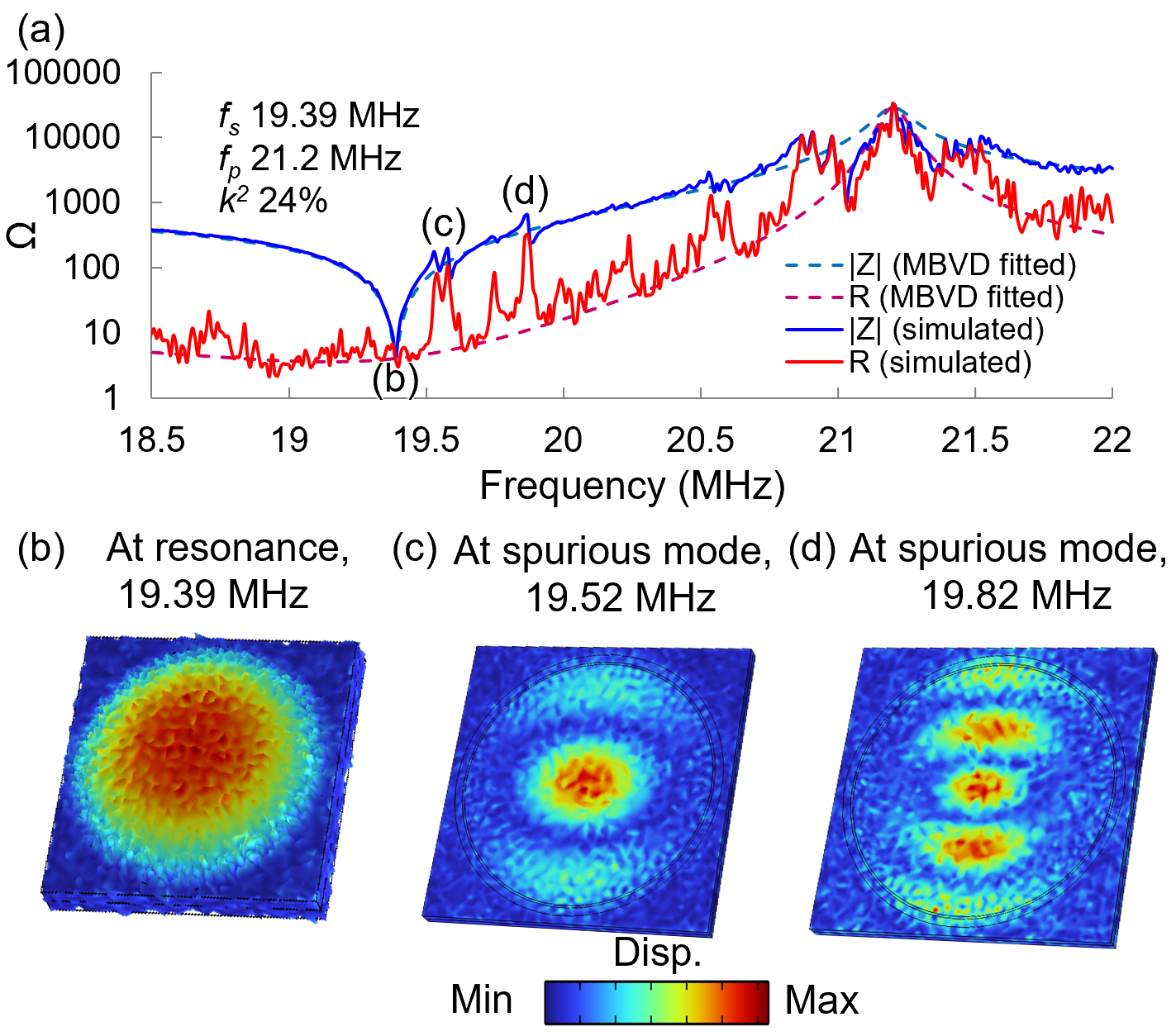}}
\caption{FEA simulated results of the proposed P3F resonator: (a) Simulated impedance and resistance showing resonance and in-band spurious modes across the full frequency range; (b) Mode shape at resonance; (c), (d) Spurious mode shapes at nearby frequencies.}
\label{fig3}
\end{figure}

Building on this mode analysis, we further evaluate the device-level performance of the proposed P3F TE-mode resonator through FEA simulations of admittance, displacement, and stress, as shown in Fig.~\ref{fig3}. The simulated admittance response Fig.~\ref{fig3}(a) shows a clear series resonance at 19.39~MHz and a parallel resonance at 21.2~MHz, yielding $k^2$ of $24\%$. The mode shape at resonance Fig.~\ref{fig3}(b) confirms a TE A2 mode with symmetric radial displacement concentrated near the center. In contrast, displacement profiles at 19.52~MHz and 19.82~MHz, Fig.~\ref{fig3}(c) and Fig.~\ref{fig3}(d), correspond to in-band spurious modes with fragmented and asymmetric distributions. Suppression of these modes will be the focus of future structural optimization, while this work demonstrates the first implementation of P3F resonators for power electronic applications.

\begin{figure}[!t]
\centerline{\includegraphics[width=\columnwidth]{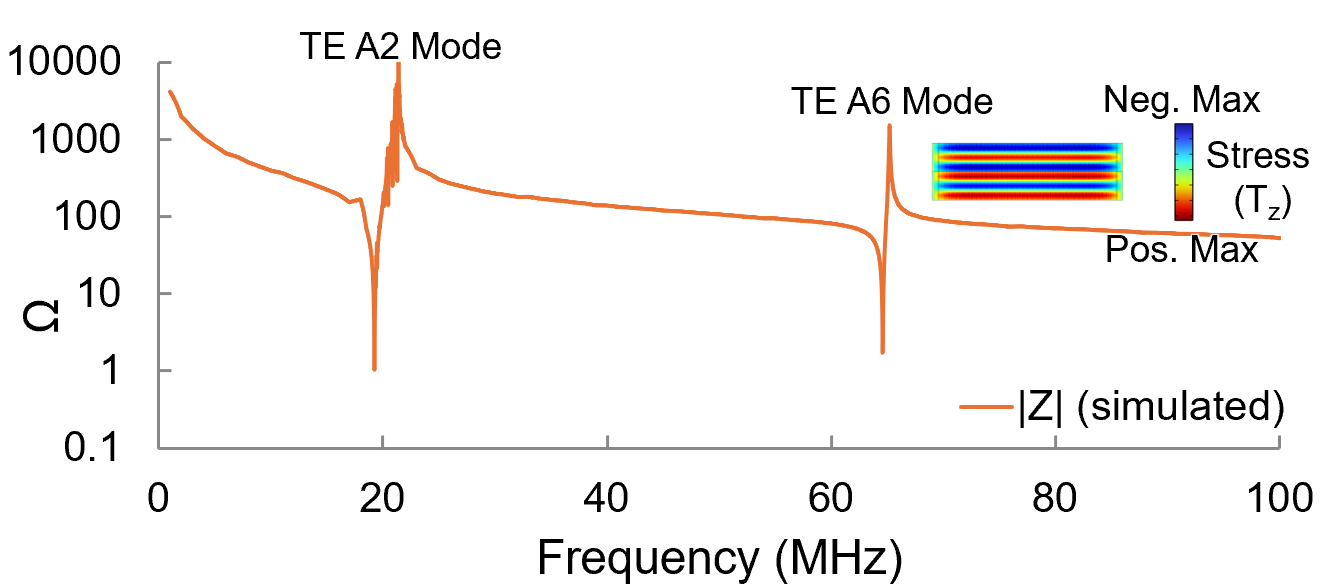}}
\caption{Simulated wideband impedance response from 1–100 MHz with cross-sectional stress profile of A6.}
\label{fig4}
\end{figure}

Figure. \ref{fig4} extends the impedance response analysis over a wide frequency band from 1 to 100~MHz. The resonator shows a 6th-order antisymmetric (A$_6$) resonance at 64.56~MHz with $k^2$ of 2.46\%. Crucially, no S$_1$ mode is excited across the entire frequency range. This is achieved by the polarity alternation in the P3F structure, where piezoelectric coefficients that are odd under inversion symmetry are inverted across layers, leading to destructive interference for odd-order modes like the S$_1$ mode, and constructive excitation of even-order TE modes such as A$_2$ and A$_6$ modes. 

\section{Wafer Bonding}

\begin{figure}[!t]
    \centering
    \includegraphics[width=\columnwidth]{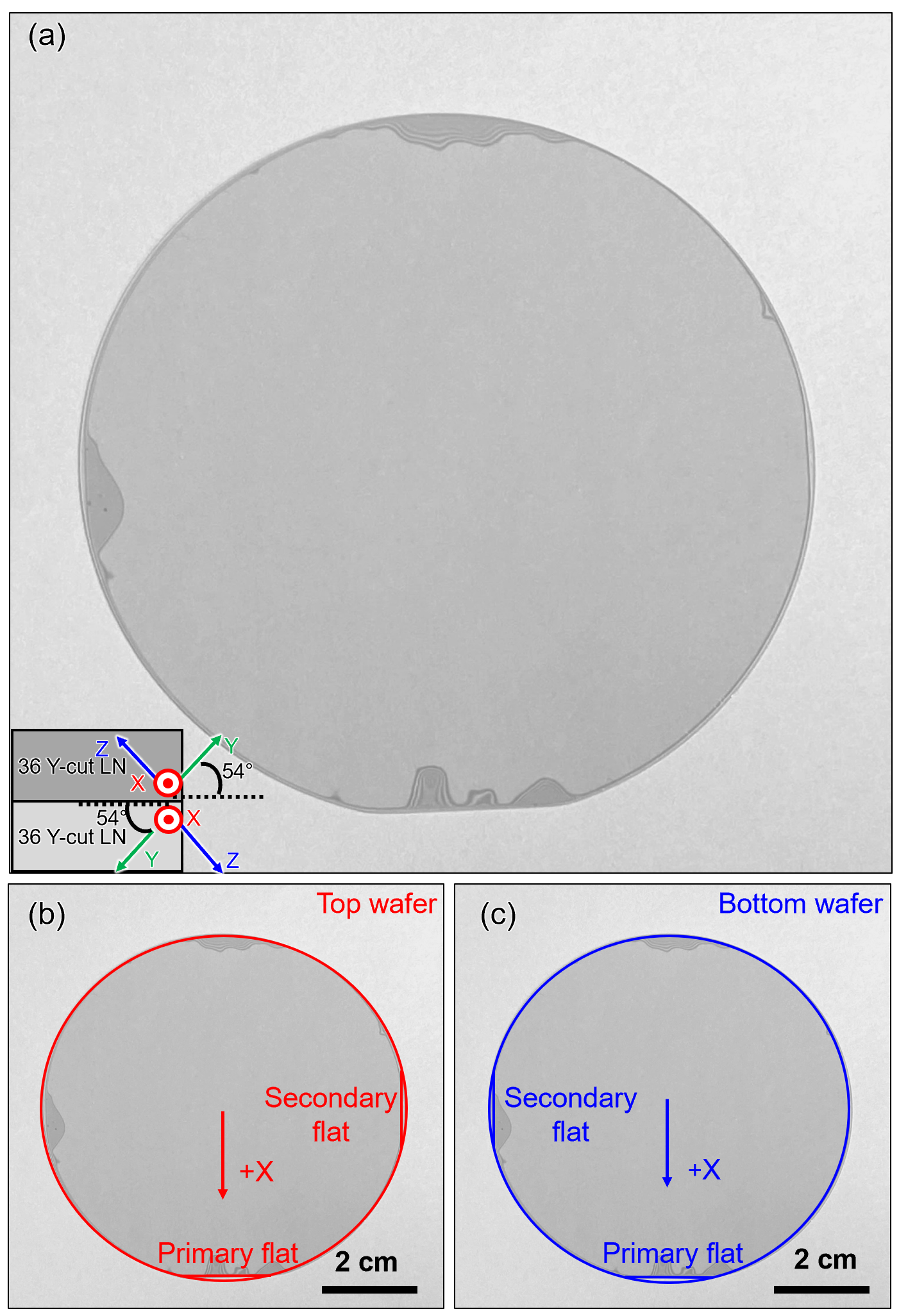}
    \caption{(a) Optical image of the bonded 36\textdegree{}Y-cut|36\textdegree{}Y-cut LiNbO\textsubscript{3} pair with 97\% effective bonded area. Inset: bonding design schematic. Schematics of wafer orientation for the (b) top and the (c) bottom wafer.}
    \label{fig:wafer_bonding}
\end{figure}

The LiNbO\textsubscript{3} structure consists of a pair of 170~µm thick 36\textdegree{} Y-cut LN wafers. Prior to bonding, the roughness of the wafer surfaces was confirmed to be less than 1 nm r.m.s. The wafers were cleaned in an SC1 solution (5:1:1 H\textsubscript{2}O:H\textsubscript{2}O\textsubscript{2}:NH\textsubscript{4}OH) for 10 min, followed by a hydrofluoric acid (HF) dip for 10 min, and then rinsed with deionized (D.I.) water. After cleaning, a surface treatment with (NH\textsubscript{4})\textsubscript{2}S was performed for 2.5 min. The wafers were rinsed again with D.I. water to remove excess solution from the surfaces and then spin-dried.

Following cleaning and surface treatment, the wafers were crystallographically aligned such that the positive 36\textdegree{} Y-cut surfaces were opposite to each other and the X-direction of both wafers was aligned identically, as illustrated in Fig.~\ref{fig:wafer_bonding}(b) and (c). After alignment, a uniform, light pressure (kPa range) was applied to initiate bonding. Subsequently, the bonded structure was annealed first at 100~\textdegree{}C for 16 hours, followed by 250~\textdegree{}C for an additional 16 hours. This resulted in greater than 97\% effective bonded area without observable voids at the interface, as shown in Fig.~\ref{fig:wafer_bonding}(a).

\begin{figure}[!t]
    \centering
    \includegraphics[width=\columnwidth]{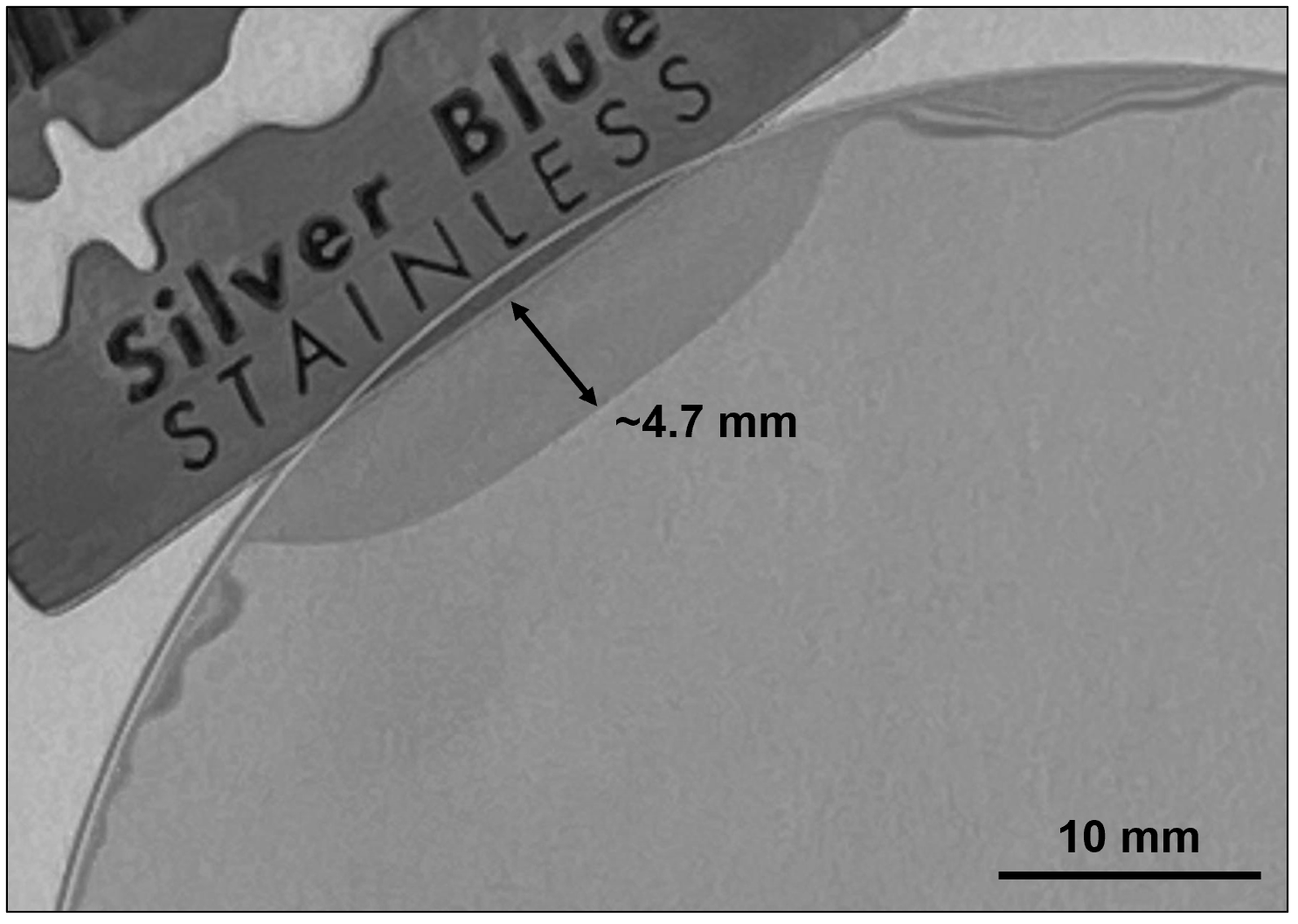}
    \caption{Bond strength measurement showing a crack length of 4.7 mm after blade insertion, corresponding to a bond strength around 1 J/m\textsuperscript{2}.}
    \label{fig:bond_strength}
\end{figure}

Bond strength measurements \cite{maszara1988bond} (Fig.~\ref{fig:bond_strength}) indicated an interfacial bond strength of approximately 1 J/m\textsuperscript{2}, comparable to previously reported values for LN|SiO\textsubscript{2} \cite{takigawa2018bond} and LN|Si \cite{takagi2001bond} bonded pairs. The bonded pair was diced into several squares for subsequent device processing.

\section{Resonator Fabrication}

\begin{figure}[!t]
\centerline{\includegraphics[width=\columnwidth]{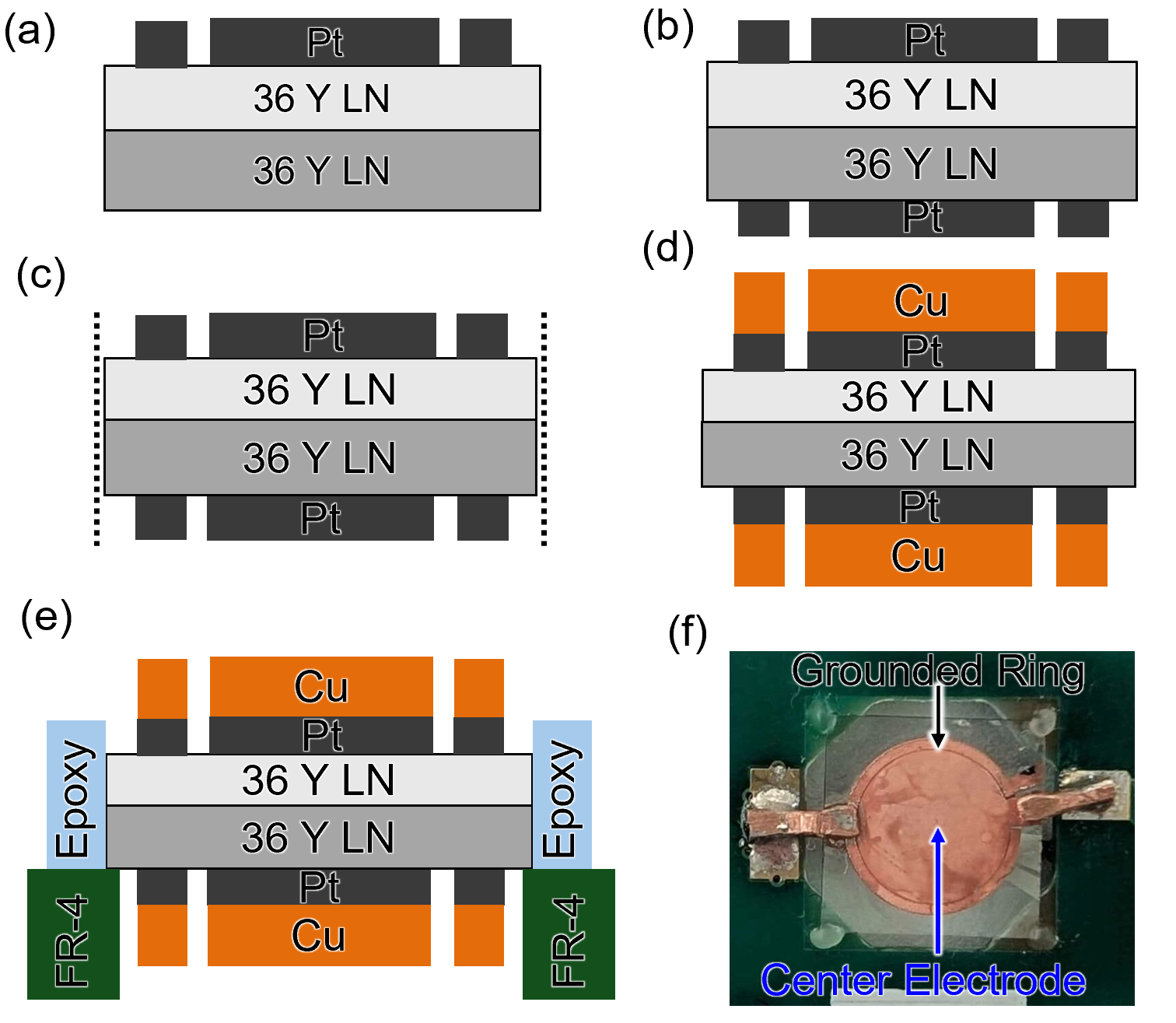}}
\caption{Fabrication and integration process for P3F TE LN resonator. 
(a) Wafer bonding of oppositely oriented 36°Y LN layers. 
(b) Top and bottom Pt electrode deposition. 
(c) Electrode patterning. 
(d) Wafer dicing and Cu pad deposition. 
(e) Final integration with thick Cu plating and epoxy bonding onto FR-4 PCB. 
(f) Fabricated resonator bonded to PCB with visible grounded ring and center electrode.}
\label{fig7}
\end{figure}

After successful wafer bonding, the P3F TE resonator is fabricated using standard cleanroom microfabrication techniques, as illustrated in Fig.~\ref{fig7}. The process begins with solvent cleaning of the bonded P3F LN wafer using acetone and isopropyl alcohol (IPA) to remove surface contaminants. A thin layer of photoresist is then spin-coated onto the top surface, and electrode patterns are defined through photolithography. Following photoresist development, a 50 nm platinum (Pt) thin-film metal layer is deposited by electron beam evaporation. Pt is selected as the electrode material due to its high thermal stability, chemical inertness, good electrical conductivity, and strong adhesion to the LN substrate. Lift-off in an acetone bath removes the remaining photoresist and excess metal, leaving well-defined top electrodes.

Next, the backside electrode structures are fabricated using the same procedure, this time employing backside alignment to ensure accurate pattern alignment to the top-side electrodes. Once both electrode layers are patterned, the wafer undergoes optical inspection to ensure proper electrical grounding of the surrounding ring and to avoid unintended feedthrough between isolated structures.

After verification, the wafer is diced into individual resonator dies. Each die is subsequently electroplated with a 500 nm layer of copper (Cu) to ensure uniform current distribution across the electrodes. The plating thickness is optimized to minimize series resistance while avoiding nonuniform plating that could degrade the resonator's quality factor \cite{bonavia2021augmented}. Finally, each resonator die is mounted onto an FR-4 PCB using UV-cure epoxy. This provides mechanical damping and electrical grounding through the bottom pads, while the central active electrode remains suspended in air to allow efficient TE mode excitation and minimize acoustic energy leakage.

\section{Measurement Results}
The fabricated P3F TE-mode LN resonator was characterized using a vector network analyzer (VNA). One-port $S_{11}$ measurements were performed using an SMA connector on the testbed, with an input power level of 10 dBm. The scattering parameters were then converted to impedance ($Z$) and resistance ($R$) for further analysis.

\begin{figure}[!t]
\centerline{\includegraphics[width=\columnwidth]{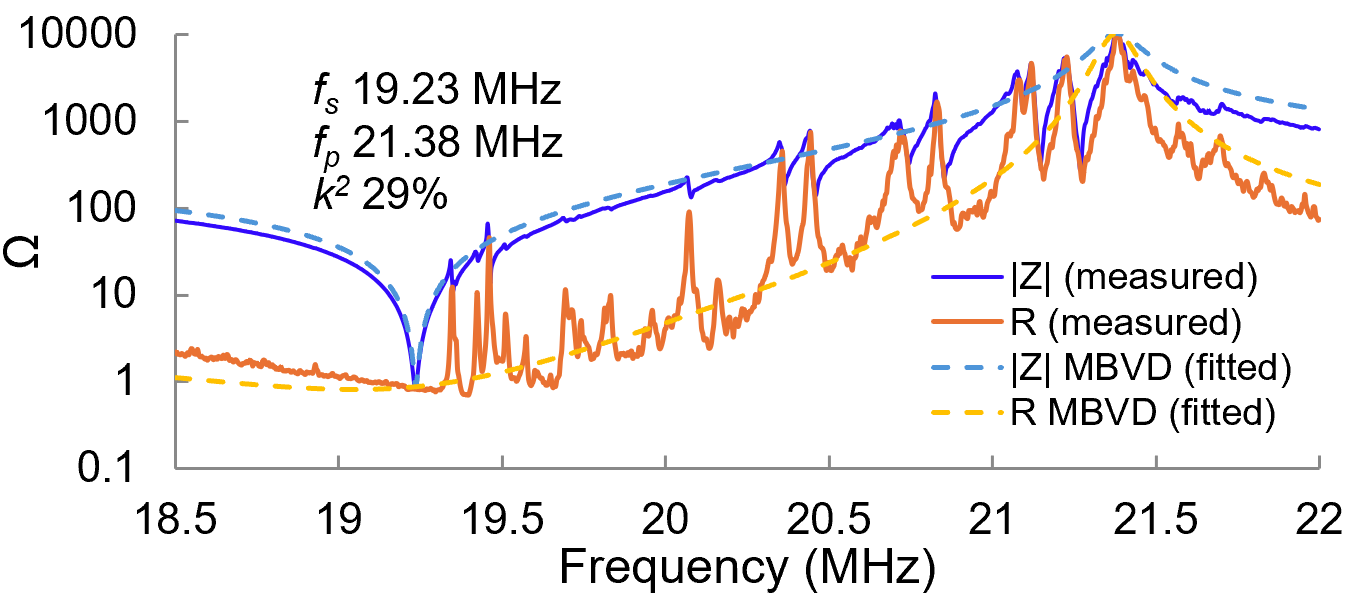}}
\caption{Measured frequency domain impedance and resistance of fabricated P3F TE LN resonator, featuring a resonance frequency at 19.23 MHz and $k^2$ of 29\%.}
\label{fig22}
\end{figure}

\begin{figure}[!t]
\centerline{\includegraphics[width=\columnwidth]{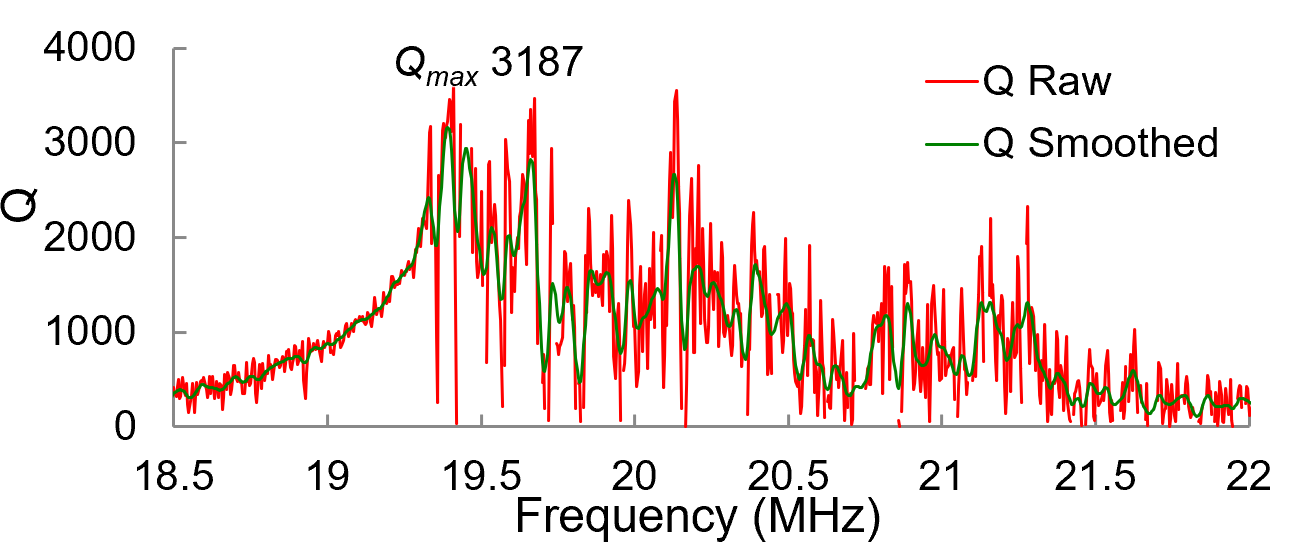}}
\caption{Raw Q and smoothed Bode Q of the P3F LN resonator.}
\label{fig11}
\end{figure}

\begin{figure}[!t]
\centerline{\includegraphics[width=\columnwidth]{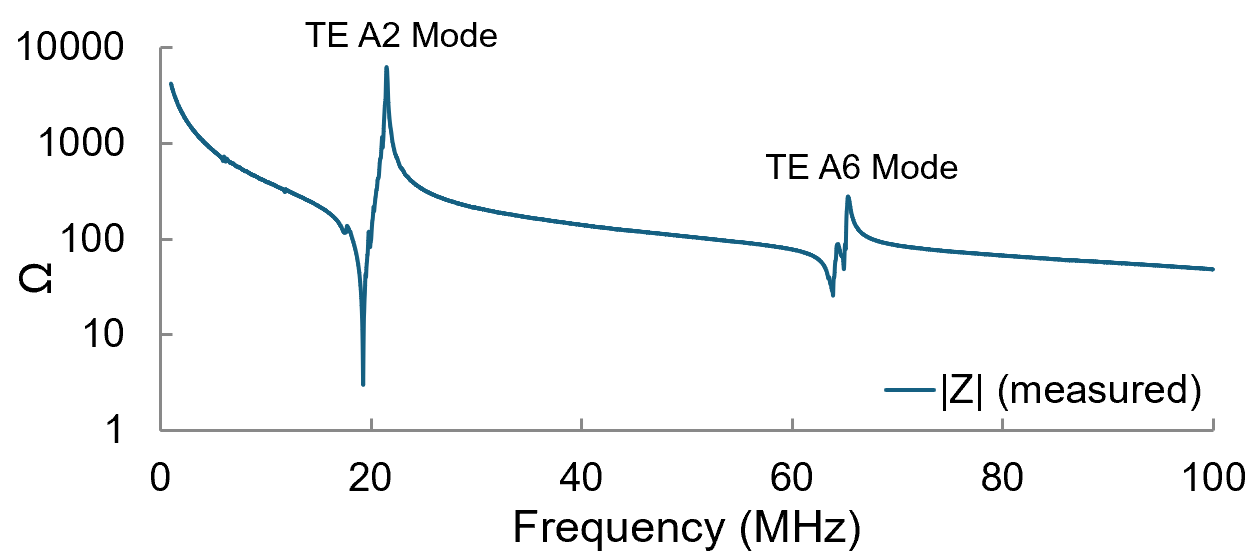}}
\caption{Measured wideband impedance response from 1–100 MHz}
\label{fig31}
\end{figure}

As shown in Fig.~\ref{fig22}, the impedance magnitude exhibits a clear minimum at $f_s = 19.23$~MHz and a maximum at $f_p = 21.38$~MHz, yielding an extracted coupling coefficient of $k^2 = 29\%$. The resistance curve confirms the resonance behavior with a sharp dip near $f_s$, while the impedance trace reveals multiple in-band peaks due to higher-order spurious modes, likely arising from lateral or Lamb wave excitations.

Fig.~\ref{fig11} shows the Bode $Q$~\cite{feld2008after} of the measured P3F LN resonator, extracted from the phase derivative of the measured S-parameters. Because Bode $Q$ is derived from group delay, it is highly sensitive to noise and spurious modes, which manifest as rapid oscillations and sharp peaks in the raw $Q$ curve. To mitigate these artifacts and better visualize the resonator's intrinsic performance, we applied a Gaussian moving average filter to the raw data. This smoothing step is crucial for accurately interpreting the broadband $Q$ trend and suppressing fluctuations resulting from lateral mode coupling or measurement imperfections. After smoothing, the device exhibits a high quality factor, with a peak value of $Q$ of 3187 near resonance. However, the 3dB $Q$ of the same device is lower at 1282, likely due to electrical loading effects from the Cu electrodes. This degradation can be mitigated by increasing the metal thickness during the electroplating process, although excessive or nonuniform plating may negatively impact the quality factor near resonance\cite{bonavia2021augmented}.

Figure~\ref{fig31} shows the measured wideband impedance response of the P3F LN resonator from 1 to 100~MHz. Two prominent resonance peaks are observed: a TE A$_2$ mode at 19.23~MHz and a higher-order TE A$_6$ mode at 63.9~MHz. Notably, the measured spectrum shows no presence of the S$_1$ mode below 10 MHz, consistent with simulation results and confirming that the P3F structure effectively cancels odd-order symmetric modes through polarity alternation. 

\begin{figure}[!t]
\centerline{\includegraphics[width=\columnwidth]{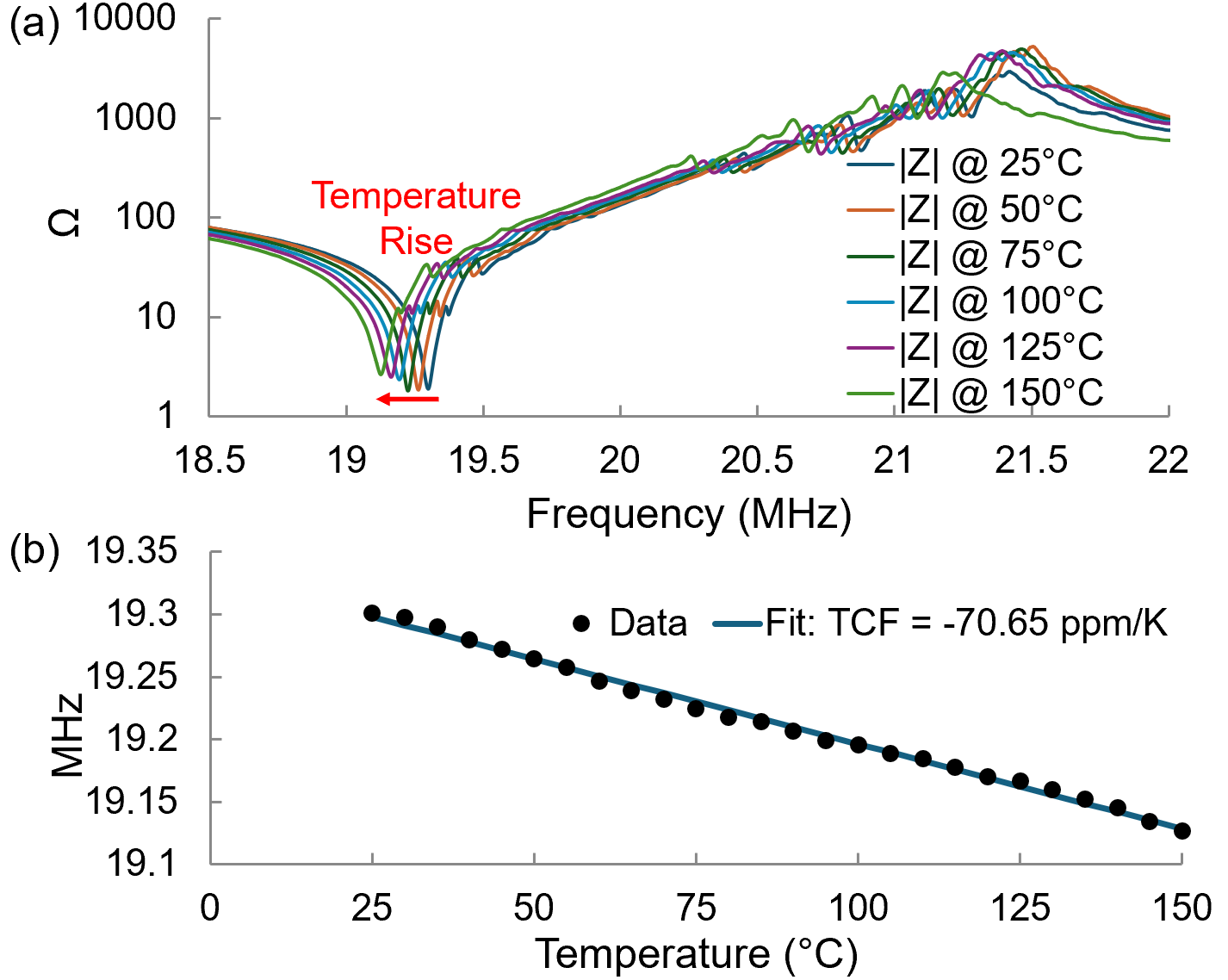}}
\caption{(a) Measured frequency domain impedance and resistance at different temperatures, along with (b) extracted TCF between 25\textdegree{}C to 150\textdegree{}C, showing a TCF of -70.65 ppm/K.}
\label{fig12}
\end{figure}

Power converters and their components often operate under a broad range of operating temperatures, making thermal stability a critical design consideration for piezoelectric resonators. To ensure reliable device performance and prevent reduced coupling \cite{miclea2006temperature}, it is crucial to investigate how temperature variations affect both the material properties and the electrical response of the device. 

To evaluate thermal stability, the LN resonator was tested over a temperature range from 25\textdegree{}C to 150\textdegree{}C. As shown in Fig.~\ref{fig11}(a), the impedance shifts progressively to lower frequencies as temperature increases, indicating a systematic reduction in $f_s$. This behavior results from the temperature-induced decrease in the Young's modulus of the LN substrate, which softens the material and lowers the acoustic velocity \cite{smith1971temperature}.

\begin{figure}[!t]
\centerline{\includegraphics[width=\columnwidth]{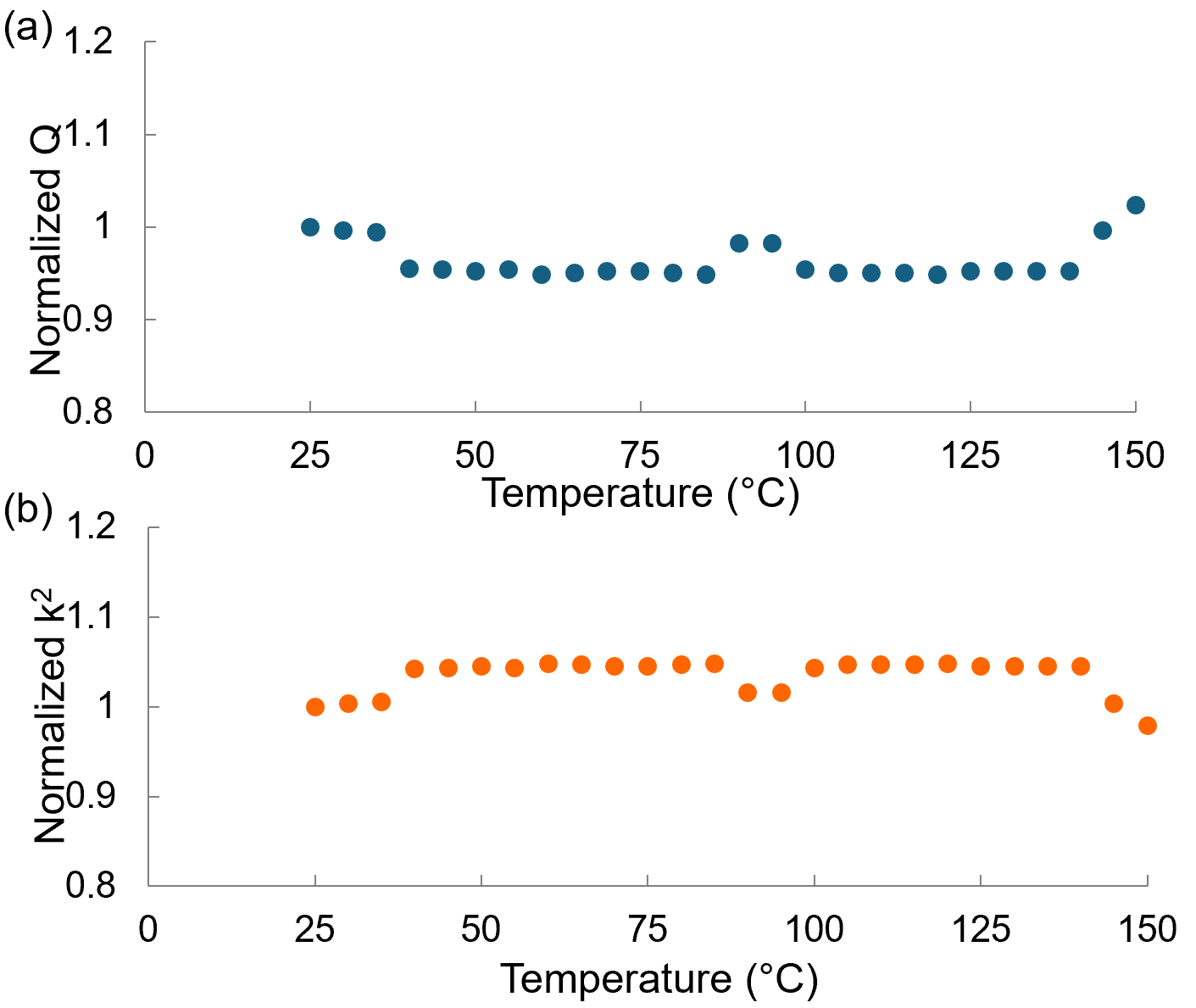}}
\caption{(a) $k^2$ and (b) $Q$ at different temperatures, extracted from the frequency domain measurement.}
\label{fig13}
\end{figure}

In addition to the frequency shift, temperature-dependent variations in device performance were also observed in the normalized $k^2$ and $Q$, fitted from the measurement in Fig.~\ref{fig12}, both normalized to their values at 25\textdegree{}C. As shown in Fig.~\ref{fig13}(a) and Fig.~\ref{fig13}(b), both the normalized $Q$ and the normalized $k^2$ remain relatively stable over most of the temperature range, with slight fluctuations attributed to minor in-band spurious modes as well as the increased acoustic and dielectric losses at elevated temperatures. These results suggest that while energy dissipation increases with temperature, the underlying coupling mechanism remains largely unaffected.

The extracted $f_s$, plotted in Fig.~\ref{fig11}(b), shows a linear dependence on temperature, with a measured temperature coefficient of frequency (TCF) of approximately $-70.65~\mathrm{ppm/K}$. While the LN TCF is higher than other commonly used piezoelectric materials such as PZT ($-35~\mathrm{ppm/K}$) and LT ($-13.56~\mathrm{ppm/K}$), LN exhibits stable and reversible thermal behavior. Importantly, no irreversible degradation was observed in the device after thermal cycling up to 150\textdegree{}C. Furthermore, previous studies have demonstrated that LN can operate reliably at temperatures up to 500\textdegree{}C thanks to its high Curie temperature~\cite{eisner2020laterally,chaudhari2025thermal}. These results demonstrate the thermal robustness of LN and support its use in high-power piezoelectric power conversion applications.

The measured P3F resonator achieves the highest $f_s$ at 19.23 MHz among all devices in Table~\ref{Talble:1}, while also maintaining a high $k^2$ of 0.29 and $Q$ of 3187. Compared to PZT and radial-mode LN devices, which are limited to lower frequencies, and single-layer LN TE-mode resonators operating below 10 MHz, the P3F design extends into a higher frequency regime without sacrificing $Q$ or $k^2$. It also achieves a FoM of 928 and the highest $f_s \cdot Q$ product, demonstrating its advantage for high-frequency, high-efficiency piezoelectric power conversion.

\section{Power Handling Analysis}

\begin{figure}[!t]
\centerline{\includegraphics[width=\columnwidth]{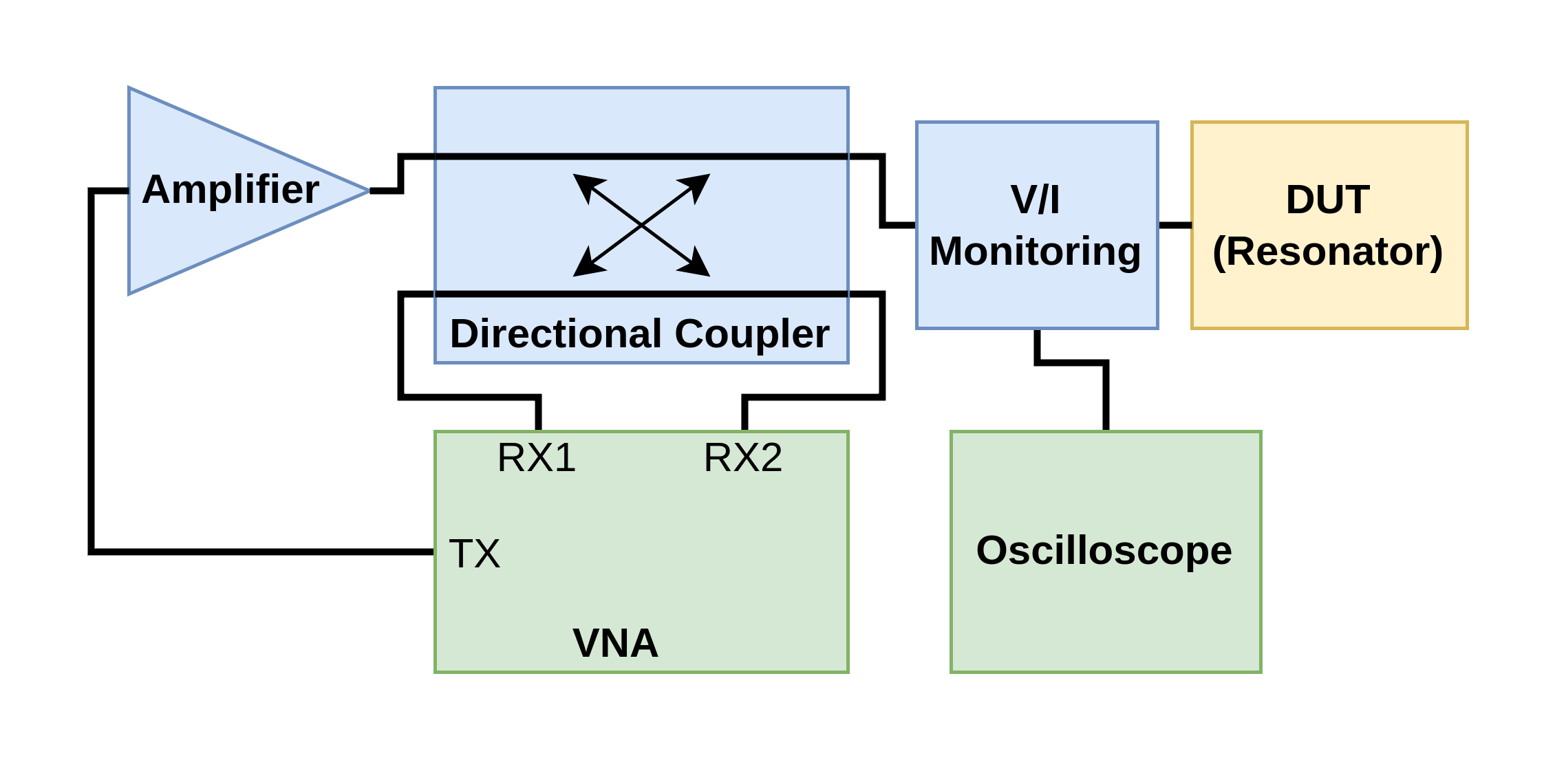}}
\caption{High power resonator impedance testing setup.}
\label{high_power_setup}
\end{figure}

The maximum power handling capability of the fabricated P3F TE-mode LN resonator can be determined experimentally through destructive electrical testing. By injecting a single-frequency sinusoidal pulse into the resonator and increasing the power level of the pulse until material failure occurs, the maximum power level that the piezoelectric material can sustain before excessive stress or strain causes it to fracture can be determined. The high power measurement is conducted using the test setup illustrated in Fig.~\ref{high_power_setup}.

A specialized VNA, the Bode 100, is used to inject a small signal into a 60~dBm power amplifier. The external directional coupler allows the Bode 100 to make an impedance measurement based on the amplified signal. This is a specialized feature of the Bode 100 vector network analyzer (VNA); other VNAs may not be designed for use with an external directional coupler, which is required to prevent damage to the VNA from the high-power signal. The amplified signal is then fed from the directional coupler into the DUT, which is monitored with voltage and current probes connected to an oscilloscope. This allows for the monitoring of voltage and current waveform amplitudes, as well as any waveform distortion that may arise. Attenuators may be added as needed, either before or after the power amplifier, to adjust signal levels. This setup is capable of driving frequency sweeps and continuous single-frequency sine waves. However, short, single-frequency pulses are preferable for this experiment as they decouple thermal effects, which can cause impedance drift as shown in Fig.~\ref{fig11} and premature material failure.  

\begin{figure}[!t]
\centerline{\includegraphics[width=\columnwidth]{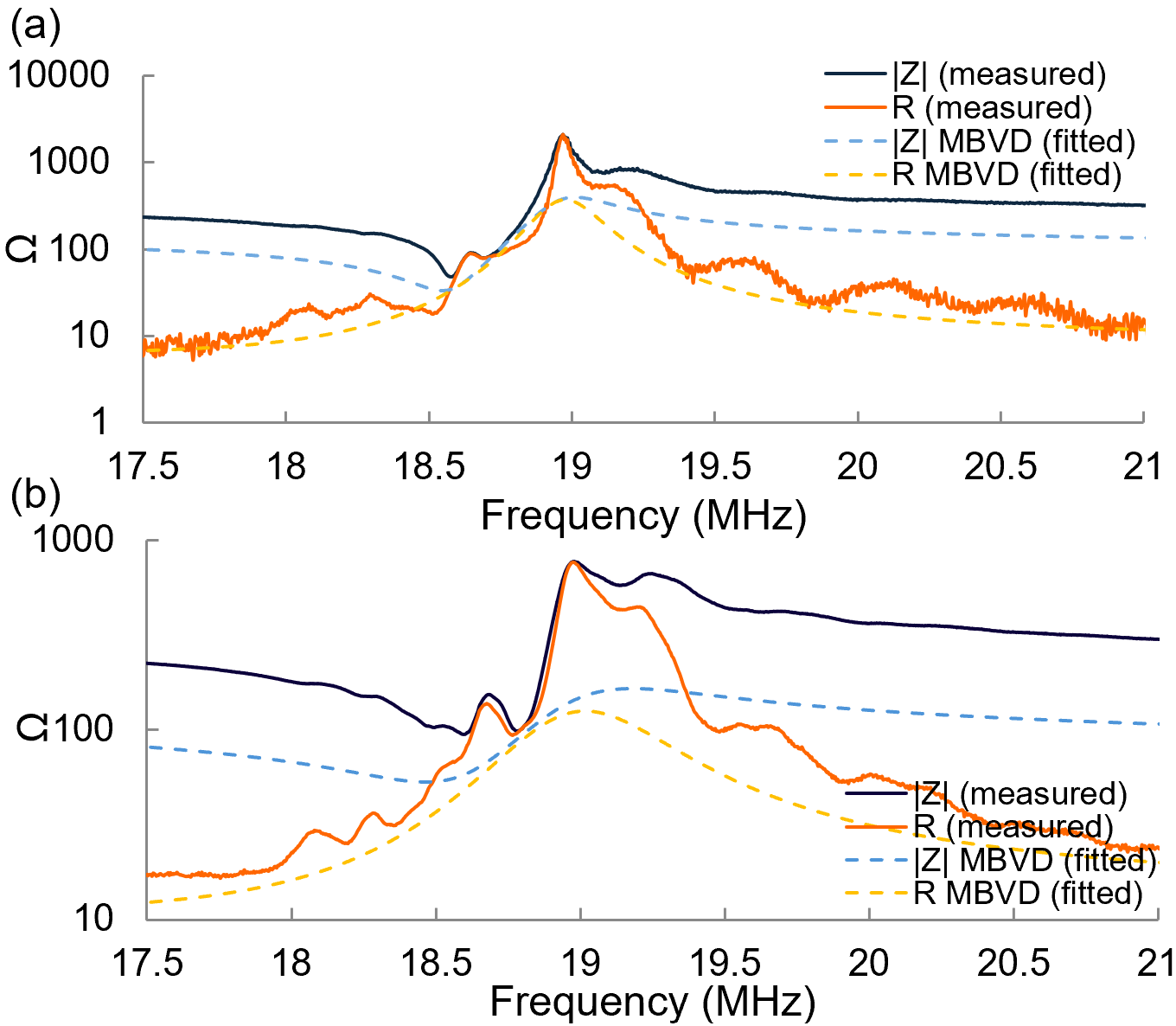}}
\caption{(a) New resonator and matching network frequency response and (b) resonator and matching network frequency response after testing at 42 dBm. Note the reduction of both the series and parallel resonant Q as well as the increased resistance.}
\label{fig8}
\end{figure}

Due to small variations in fabrication and electroplating, multiple devices of identical geometries can have slightly different frequency responses. To test the other devices at a consistent operating point, each device is matched using an added series capacitance, ensuring the matched resonant frequency corresponds to the frequency at which the unmatched device impedance is approximately 150~$\Omega$. This ensures that all devices are driven at the same impedance while bringing the impedance seen by the power amplifier closer to 50~$\Omega$. This also reduces any possible waveform distortion due to impedance mismatch. The impedance of the resonator with the matching network is pictured in Fig.~\ref{fig8}(a).

\begin{figure}[!t]
\centerline{\includegraphics[width=\columnwidth]{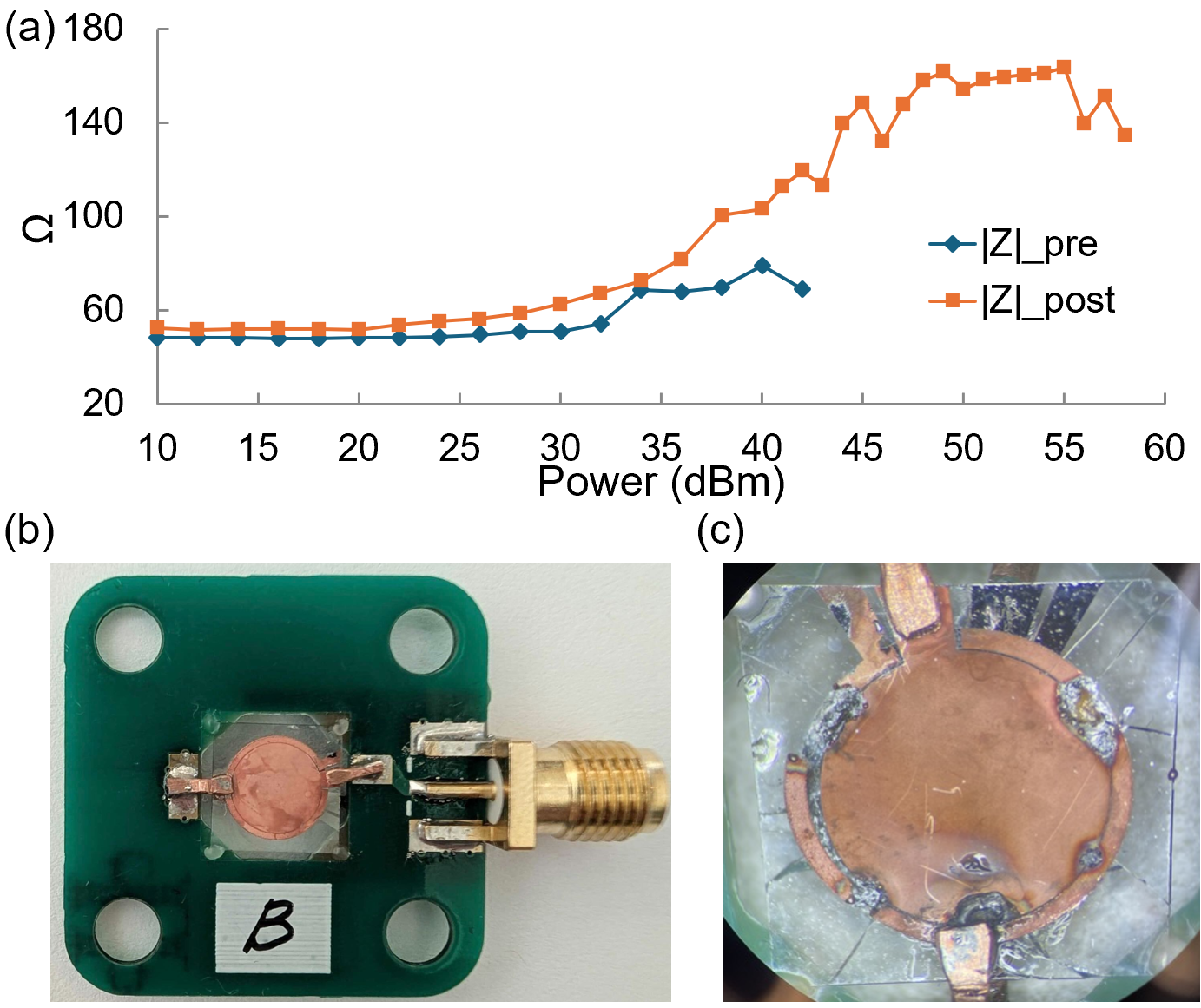}}
\caption{(a) Normalized |Z| vs power of new resonator from 10dBm to 42 dBm (blue points) at which point bonding damage is suspected, and of same, damaged resonator from 10 dBm to 58 dBm (orange points) at which point the device fully fails (b). (c) Device after PCB mounting, before power handling testing. Damaged resonator at 58 dBm with visible cracks and electrodes blackened and burned off.}
\label{fig9}
\end{figure}

A new, electroplated, and mounted device used for this testing is shown in Fig.~\ref{fig7}. Fig.~\ref{fig9}(a), blue points, shows the magnitude of the impedance of the new device normalized to its nominal small signal impedance of 48.312~$\Omega$ as the power level is increased. Between 10~dBm and 32~dBm, the impedance magnitude remains relatively constant, deviating only 12\% from the small signal value. Between 32~dBm and 42~dBm, however, the impedance magnitude deviates an additional 51\%. At 42~dBm, a permanent change in the frequency response is observed as shown in Fig.~\ref{fig8}; the altered frequency response persists after ramping down to lower power levels. This failure mode was confirmed in a second test, with a second resonator of identical geometry exhibiting the same failure mode at 38~dBm. This change is likely due to bonding damage and occurs at a lower power level than full failure of the lithium niobate itself, which will be the focus for future study.

The impedance magnitude of the damaged device normalized to its new nominal small signal value of 52.44~$\Omega$ as the power level is increased is shown in Fig.~\ref{fig9}(a). Again, the impedance level remains relatively constant at low signal power. Above 42~dBm, however, the impedance measurement becomes volatile with multiple measurements of the same device at the same frequency yielding different results. The device deviates up to 211\% from its nominal small signal impedance before fully failing at 58~dBm with a voltage of 463.3V and a current of 3.342A. As shown in Fig.~\ref{fig9}(b), cracks are visible in the lithium niobate, and the electrodes are singed or burned off in the ring area.

While this measured failure point applies only at a single frequency, more points may be tested to provide a more comprehensive understanding of the device's material limits, which will be a topic for future study.


\section{Conclusion}

This work presents a novel P3F TE-mode bi-layer LN resonator demonstrating a high quality factor ($Q$) of 3148 and a strong electromechanical coupling coefficient ($k^2$) of 29\% at 19.23~MHz. Thermal testing confirms that the device maintains robust performance across a wide temperature range, while high-power testing reveals its power handling up to a failure threshold of 58 dBm. The state-of-the-art (\textit{ $f_s \cdot Q$}) product, combined with high resilience under elevated thermal and electrical stress, demonstrates the potential of P3F LN resonators for next-generation piezoelectric power conversion. With further design optimization and packaging improvements, this platform can enable compact, efficient, and spurious-free operation in high-frequency, high-power applications.

\section*{Acknowledgment}
The authors would like to thank Dr. Todd Bauer and Dr. Sunil Bhave for the helpful discussion. 

\end{document}